\documentclass[12pt,a4paper]{article}

\usepackage[utf8]{inputenc}
\usepackage[T1]{fontenc}
\usepackage[english]{babel}
\usepackage{amsmath,amssymb}
\usepackage{graphicx}
\usepackage{booktabs}
\usepackage{array}
\usepackage{hyperref}
\usepackage{geometry}
\usepackage{caption}
\usepackage{natbib}
\usepackage{microtype}
\usepackage{xcolor}

\hypersetup{
  colorlinks=true,
  linkcolor=blue!60!black,
  citecolor=blue!60!black,
  urlcolor=blue!60!black
}

\newcommand{\Tstar}{T^*}

\title{\textbf{A 1151-Year Quasi-Commensurability \\of the Solar System: Empirical Detection, Statistical Characterization, \\
and the Anomalous Exclusion of Uranus}}

\author{Carlos \textsc{Baiget Orts}\thanks{Correspondence:
  \href{mailto:asinfreedom@gmail.com}{asinfreedom@gmail.com}.
  ORCID: \href{https://orcid.org/0009-0000-6725-5188}{0009-0000-6725-5188}.
  Family name: Baiget Orts.}\\
  \small\textit{Independent researcher, Valencia, Spain}}

\date{}

\begin{document}
\maketitle

\begin{abstract}
We report the empirical detection of a multi-planet quasi-commensurability
in the Solar System and identify an anomalous exclusion that may bear on
the dynamical history of Uranus.
An exhaustive search identifies $\Tstar = 420{,}403$~days
($\approx 1{,}151$~years) as the global minimum of a series-comparison
similarity metric applied to daily heliocentric ecliptic longitudes of
seven planets --- Mercury, Venus, Earth, Mars, Jupiter, Saturn, and
Neptune --- computed from the DE441 ephemeris over $\pm 1{,}300$~years.
At this interval, the mean simultaneous angular displacement of all seven
planets is $13.4^\circ$, with a standard deviation of $0.65^\circ$
sustained over a century-long window and stable across $1{,}200$~years
of reference epochs.
$\Tstar$ ranks first among all $2{,}600$ candidates, with a gap of 
$1.09^\circ$ to the second best.
No sub-multiple produces a comparable result.
Seven of the eight planets participate in the synchronism.
The sole exception is Uranus, whose sidereal residue at $\Tstar$
is $-108.3^\circ$ --- nearly one-third of a full orbit --- while Neptune’s residue is only $-5.2^\circ$, making it the second smallest after Earth’s.
This sharp asymmetry between the two ice giants constitutes an independent
empirical signature consistent with the hypothesis that Uranus's orbital
period was substantially modified by a catastrophic early impact.
The interval $1{,}151$~years was identified by Babylonian astronomers as
the Venus return period \citep{dejong2019}; the present work shows it is
simultaneously optimal for six additional planets.
Source code and data are publicly available.
\end{abstract}

\clearpage
\tableofcontents
\newpage

\section{Introduction}
\label{sec:intro}

Commensurabilities between planetary orbital periods have been studied
since antiquity and play a central role in modern celestial mechanics.
Well-known examples include the Saros cycle ($\approx 18$~yr,
eclipses), the Metonic cycle ($19$~yr, Sun--Moon), and the
Jupiter--Saturn Great Conjunction cycle ($\approx 19.86$~yr).
These are all two-body near-commensurabilities, involving at most two
or three bodies simultaneously.

A fundamentally different question is whether a single time interval
can act as an approximate common multiple of the orbital periods of
most or all planets simultaneously, producing a near-simultaneous
return of the full planetary configuration.
This multi-body problem has not, to our knowledge, been addressed
systematically in the modern literature.
Existing studies of planetary commensurability typically treat pairs or
triples of bodies \citep{shirley1997_commensurability}, or investigate
analytical relations among mean orbital periods \citep{jelbring2013},
but do not perform an exhaustive empirical search for a common
quasi-period of all planets using modern high-precision ephemerides.

The derived interval of $1{,}151$~years has an independent historical significance.
Babylonian astronomers determined that Venus returns to the same
position in the sky after $1{,}151$~years.
Specifically, both System~A$_0$ and System~A$_3$ of Babylonian Venus
theory rest on the relation: in $1{,}151$~years, Venus completes $720$
synodic cycles ($= 1{,}871$ sidereal orbits), returning to precisely
the same ecliptic longitude \citep{dejong2019}.
Whether this same interval constitutes an optimal simultaneous
quasi-period for the remaining planets has never been investigated.

This paper answers the multi-body commensurability question through a
fully computational, reproducible approach.
We define a rigorous similarity metric based on series comparison,
apply it exhaustively over a $\pm 1{,}300$-year search range using
the DE441 ephemeris, and characterise the resulting global minimum
statistically.
The analysis has two objectives: (i)~to demonstrate the existence and
robustness of the 1{,}151-year planetary quasi-commensurability; and
(ii)~to show that Uranus is the sole planet excluded from it, and to
discuss this exclusion as an independent empirical line of evidence
consistent with a catastrophic perturbation of Uranus's orbit \citep{kegerreis2018}.

The main findings are:

\begin{enumerate}
  \item $\Tstar = 420{,}403$~days ($\approx 1{,}151$~years) is the
    global minimum of the similarity metric over $2{,}600$ candidate
    intervals, for seven planets (Mercury through Saturn plus Neptune).
\item The score of $\Tstar$ ($14.04^\circ$) is the lowest of all 
    $2{,}600$ candidates (rank $1/2{,}600$), with a gap of $1.09^\circ$ 
    to the second best.
  \item The result is stable: the score varies by less than $0.11^\circ$
    over any reference epoch in a $1{,}200$-year window.
  \item $\Tstar$ is the global minimum for series as short as one year,
    with a stable gap of $\approx 1.1^\circ$ to the second-best
    candidate.
  \item $\Tstar$ is irreducible: sub-multiples produce significantly
    worse scores.
  \item Neptune's sidereal residue at $\Tstar$ is $-5.2^\circ$, the
    result of the near-integer relation
    $\Tstar \approx 7 \times P_\text{Neptune}$.
    Uranus's residue is $-108.3^\circ$; it is the sole non-participant.
\end{enumerate}

Section~\ref{sec:method} describes the metric and procedure.
Section~\ref{sec:results} presents all numerical results.
Section~\ref{sec:discussion} discusses the arithmetic interpretation,
geocentric consequences including retrograde synchronisation, the
Babylonian Venus period, the effect of the Moon, and the significance
of Uranus's exclusion.
Section~\ref{sec:conclusions} summarises the conclusions.

\section{Method}
\label{sec:method}

\subsection{Heliocentric ecliptic positions}

All planetary positions are computed as heliocentric ecliptic
longitudes in the J2000 reference frame, using the Skyfield astronomy
library \citep{skyfield} with the DE441 ephemeris \citep{park2021}.
The seven bodies included in the primary analysis are listed in
Table~\ref{tab:bodies}.
The Earth is treated as the Earth--Moon system barycentre; for the
other planets, system barycentres are used.

\begin{table}[h]
\centering
\caption{Seven planets included in the primary analysis.}
\label{tab:bodies}
\begin{tabular}{lrr}
\toprule
Planet  & Sidereal period (days) & Sidereal period (yr) \\
\midrule
Mercury &            87.969 &   0.241 \\
Venus   &           224.701 &   0.615 \\
Earth   &           365.250 &   1.000 \\
Mars    &           686.971 &   1.881 \\
Jupiter & $4{,}332.589$     &  11.862 \\
Saturn  & $10{,}759.220$    &  29.457 \\
Neptune & $60{,}182.0$      & 164.791 \\
\bottomrule
\end{tabular}
\end{table}

The primary analysis was initially performed on six planets (Mercury, 
Venus, Earth, Mars, Jupiter, Saturn). Neptune was subsequently included after 
its sidereal residue at $\Tstar$ was found to be only $-5.2^\circ$, 
confirming its participation. Uranus was tested and found not to 
participate (residue $-108.3^\circ$); it is therefore reported 
separately as an empirical finding in 
Section~\ref{sec:results:residues} and discussed in 
Section~\ref{sec:discussion:uranus}. 

\subsection{Similarity metric}

Let $\lambda_k(t)$ be the heliocentric ecliptic longitude of planet $k$
on day $t$.
For a candidate offset $T$ (days), reference epoch $t_0$, and series
length $N$ (days), define the daily mean angular displacement:

\begin{equation}
  \delta_i(T) = \frac{1}{7}\sum_{k=1}^{7}
  d\!\left(\lambda_k(t_0+i),\;\lambda_k(t_0-T+i)\right),
  \quad i=0,\ldots,N-1,
\end{equation}

where $d(\alpha,\beta) = \min(|\alpha-\beta|\bmod 360^\circ,\;
360^\circ - |\alpha-\beta|\bmod 360^\circ)$ is the circular angular
distance.
The scalar score is

\begin{equation}
  S(T) = \overline{\delta}(T) + \sigma_\delta(T),
  \label{eq:score}
\end{equation}

where $\overline{\delta}$ and $\sigma_\delta$ are the mean and standard
deviation of $\{\delta_i(T)\}$.

The mean $\overline{\delta}$ measures average positional proximity.
The standard deviation $\sigma_\delta$ penalises temporal instability:
a low $\sigma_\delta$ means the offset between the two configurations
remains nearly constant throughout the series.

Since $\sigma_\delta \approx 0.65^\circ$ over any series length from 
one year upwards (Table~\ref{tab:convergence}), and the score remains 
stable across a complete cycle of $\Tstar$ (Table~\ref{tab:stability}), 
the offset between the two configurations is sustained dynamically 
over centuries. The geocentric consequences of this stability are 
discussed in Section~\ref{sec:discussion:retrogrades}.

The equal weighting of all seven planets reflects the absence of any
a priori reason to privilege one planet over another in a search for
a global quasi-period; the metric is deliberately planet-agnostic.
The robustness of $\Tstar$ across all reference epochs
(Section~\ref{sec:results:stability}) and all series lengths
(Section~\ref{sec:results:convergence}) demonstrates that the global
minimum is not an artefact of the metric's specific form.

\subsection{Computational procedure}

Daily positions for all seven planets are precomputed and cached for
the full date range.
The reference epoch is 15~June~0~CE (JD~$1{,}721{,}224$), using the
astronomical year convention in which year~0 corresponds to 1~BCE;
this choice is arbitrary and does not affect the result, as
demonstrated in Section~\ref{sec:results:stability}.

Candidates: $y \in [-1{,}300, +1{,}300]$~yr relative to the 
reference, step $1$~yr ($2{,}601$ candidates, including the 
self-comparison at $\Delta t = 0$ which is excluded from the 
statistical analysis). This range slightly exceeds one full cycle 
of $\Tstar$, which is sufficient since any quasi-period longer than 
$\Tstar$ would necessarily be a multiple of it. Series length 
$N = 36{,}525$~days (100 Julian years).

All computations use Python with NumPy \citep{numpy}.
Source code is publicly available \citep{code2025}.

\section{Results}
\label{sec:results}

\subsection{Global minimum}
\label{sec:results:global}

\begin{figure}[h]
  \centering
  \includegraphics[width=0.85\textwidth]{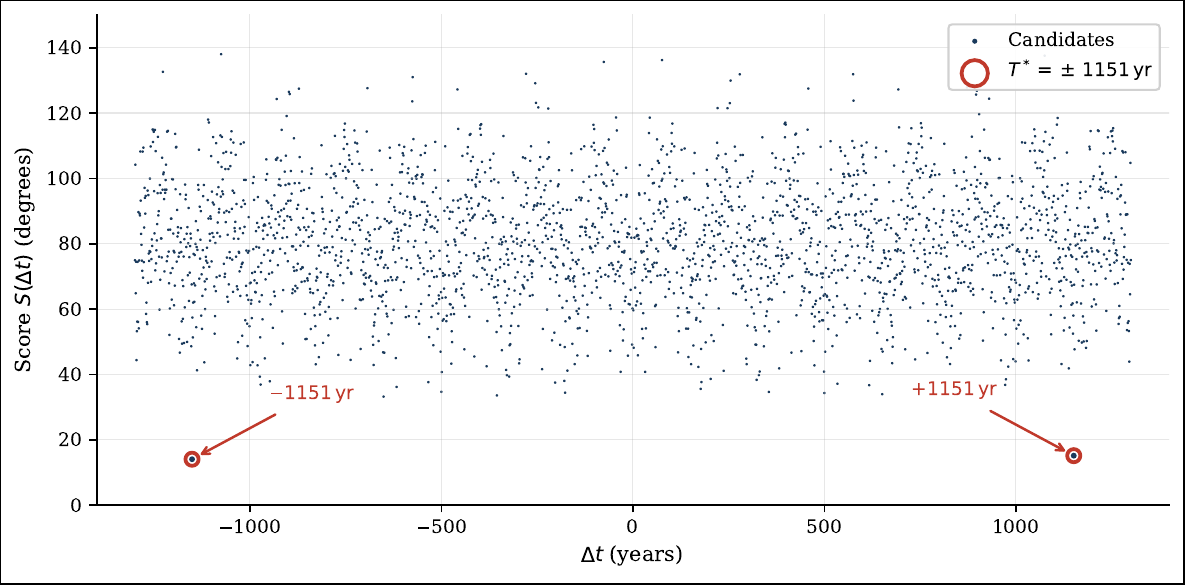}
  \caption{Score $S(T)$ for all candidate intervals in
    $[-1{,}300, +1{,}300]$~years.
    The vast majority of candidates cluster between $50^\circ$ and
    $140^\circ$.
    The two candidates at $\pm 1{,}151$~yr stand far below all others.}
  \label{fig:scatter}
\end{figure}

Figure~\ref{fig:scatter} shows $S(T)$ for all candidates.
The global minimum is

\begin{equation}
  \Tstar = 420{,}403~\text{days} \approx 1{,}151.001~\text{years},
\end{equation}

with $S(\Tstar) = 14.04^\circ$
($\overline{\delta} = 13.39^\circ$, $\sigma_\delta = 0.65^\circ$).
The second-best candidate is $+420{,}403$~days with score $15.13^\circ$.

\subsection{Statistical significance}

\begin{figure}[h]
  \centering
  \includegraphics[width=0.85\textwidth]{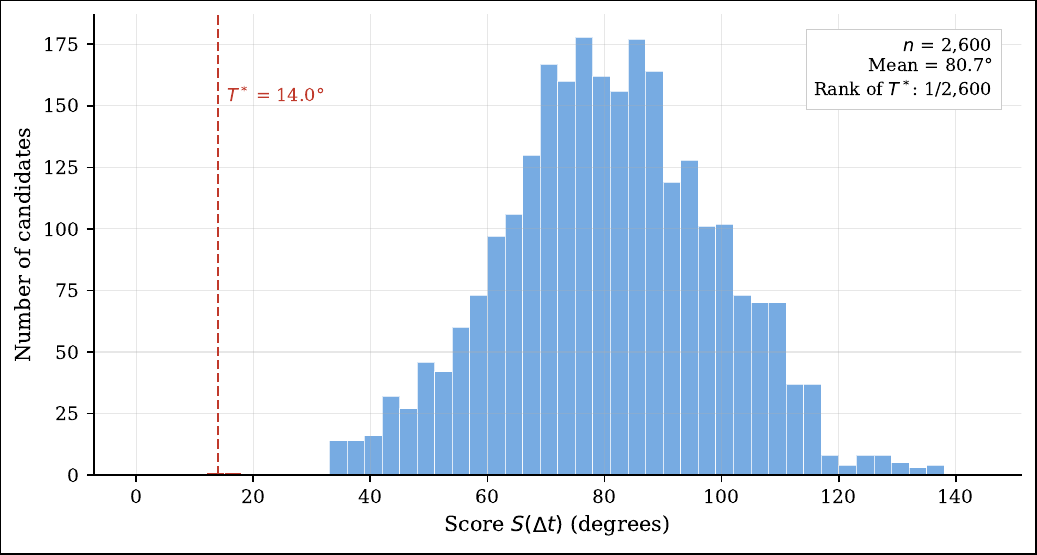}
  \caption{Score distribution.
    The single candidate with $S < 15^\circ$ is $\Tstar$;
    no other falls below $15.1^\circ$.}
  \label{fig:histogram}
\end{figure}

Table~\ref{tab:stats} summarises the score distribution.
The score distribution (Figure~\ref{fig:histogram}) is approximately 
bell-shaped, centered near $80^\circ$.
$\Tstar$ produces the lowest score of all $2{,}600$ non-zero candidates 
(rank $1/2{,}600$), with a gap of $1.09^\circ$ to the second-best 
candidate --- a gap that remains stable across all series lengths tested 
(Table~\ref{tab:convergence}).
\begin{table}[h]
\centering
\caption{Statistical characterization of the score distribution
  ($n = 2{,}600$ non-zero candidates, 7-planet analysis).}
\label{tab:stats}
\begin{tabular}{lr}
\toprule
Statistic & Value \\
\midrule
Mean score             & $80.66^\circ$ \\
Std score              & $18.22^\circ$ \\
Median score           & $80.53^\circ$ \\
Best score ($\Tstar$)  & $14.04^\circ$ \\
Maximum score          & $138.00^\circ$ \\
\midrule
Rank of $\Tstar$       & $1 / 2{,}600$ \\
Gap to 2nd best        & $1.09^\circ$ \\
\bottomrule
\end{tabular}
\end{table}

\clearpage
\subsection{Temporal stability}
\label{sec:results:stability}

Table~\ref{tab:stability} shows $S(\Tstar)$ for 12 reference epochs 
spanning $1{,}210$~years --- a range slightly exceeding one full 
cycle of $\Tstar$ --- confirming that the result is independent of 
the choice of reference epoch within a complete cycle.

\begin{table}[h]
\centering
\caption{Score $S(\Tstar)$ for 12 reference epochs.}
\label{tab:stability}
\begin{tabular}{rrrr}
\toprule
Reference (CE) & $\overline{\delta}$ ($^\circ$) &
  $\sigma_\delta$ ($^\circ$) & Score ($^\circ$) \\
\midrule
$-100$ & 13.358 & 0.640 & 13.998 \\
$+10$  & 13.358 & 0.650 & 14.008 \\
$+120$ & 13.402 & 0.653 & 14.055 \\
$+230$ & 13.462 & 0.653 & 14.115 \\
$+340$ & 13.362 & 0.642 & 14.005 \\
$+450$ & 13.230 & 0.635 & 13.865 \\
$+560$ & 13.208 & 0.630 & 13.838 \\
$+670$ & 13.197 & 0.630 & 13.827 \\
$+780$ & 13.165 & 0.631 & 13.795 \\
$+890$ & 13.210 & 0.642 & 13.853 \\
$+1000$ & 13.326 & 0.650 & 13.977 \\
$+1100$ & 13.361 & 0.652 & 14.013 \\
\midrule
Range & \multicolumn{3}{c}{$13.795^\circ$--$14.115^\circ$} \\
Std   & \multicolumn{3}{c}{$0.100^\circ$} \\
\bottomrule
\end{tabular}
\end{table}

\subsection{Angular offset time series}
\label{sec:results:timeseries}

\begin{figure}[h]
  \centering
  \includegraphics[width=0.92\textwidth]{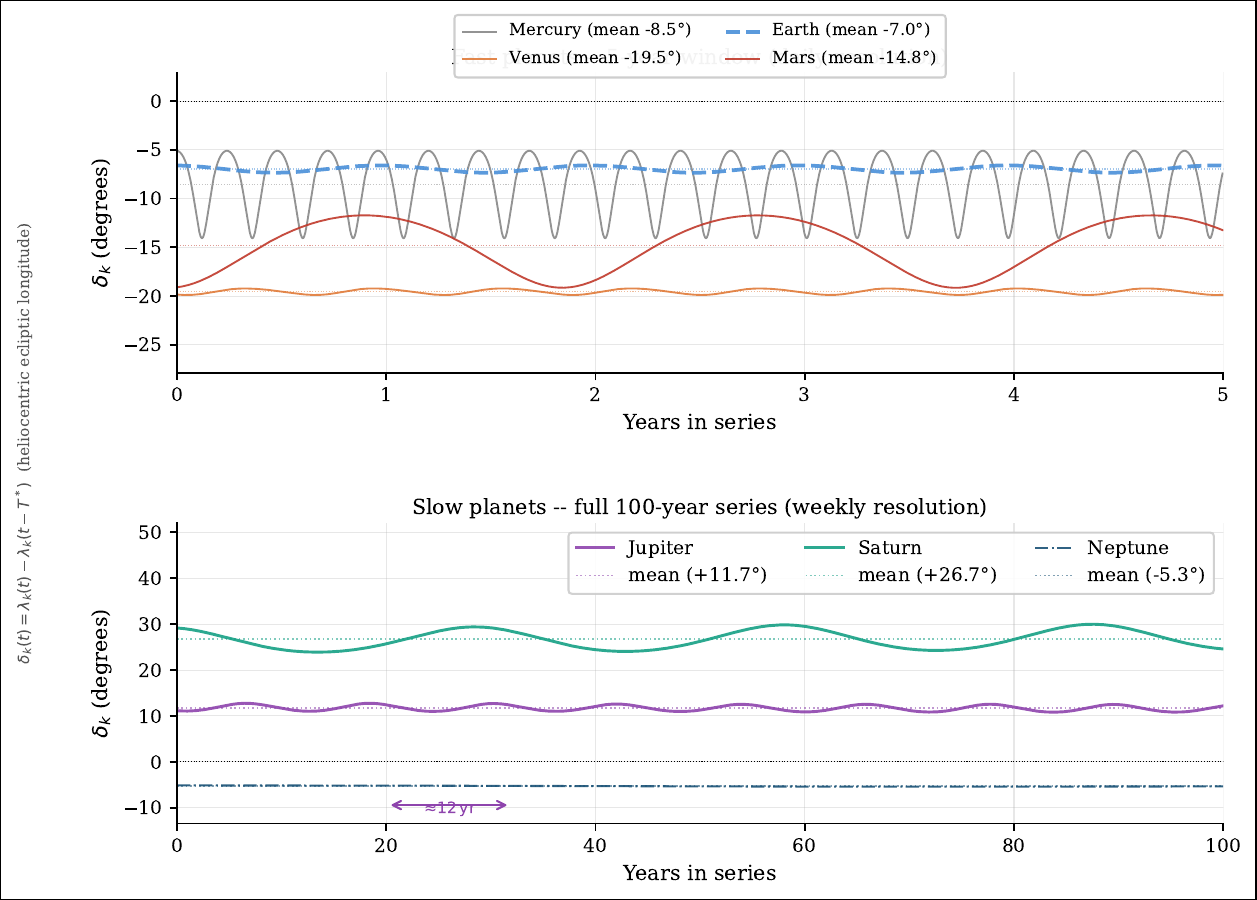}
  \caption{Daily angular offset $\delta_k(t) = \lambda_k(t) -
    \lambda_k(t - \Tstar)$ for each planet over the 100-year
    comparison series.
    \emph{Upper panel}: Mercury, Venus, Earth, and Mars over 5~years
    (daily resolution). Dashed lines show the mean offset for each
    planet.
    \emph{Lower panel}: Jupiter, Saturn, and Neptune over 100~years
    (weekly resolution). Neptune's offset is nearly constant
    ($\sigma = 0.10^\circ$), consistent with its small sidereal
    residue of $-5.2^\circ$.
    The annotated double arrow indicates the $\approx 12$-year
    Jupiter--Saturn conjunction period visible as a modulation in
    Jupiter's offset.
    Each planet oscillates around a stable mean, demonstrating that
    the quasi-commensurability is sustained dynamically over
    centuries.}
  \label{fig:panels}
\end{figure}

Figure~\ref{fig:panels} shows the daily angular offset $\delta_k(t)$
for each planet over the 100-year comparison window at $\Tstar$.
The upper panel shows the four fast planets (Mercury, Venus, Earth,
Mars) over a 5-year window at daily resolution; the lower panel shows
Jupiter, Saturn, and Neptune over the full 100-year series at weekly
resolution.
Each planet oscillates around a nearly constant mean value throughout
the series, demonstrating that the offset is sustained and not merely
a transient coincidence.
Neptune's curve is particularly flat ($\sigma = 0.10^\circ$, the
lowest of all planets), reflecting the precision of the near-integer
relation $\Tstar \approx 7 \times P_\text{Neptune}$.
The standard deviation of the daily mean across all seven planets is
$0.65^\circ$ (Table~\ref{tab:stability}), confirming this stability.

\subsection{Convergence with series length}
\label{sec:results:convergence}

\begin{figure}[h]
  \centering
  \includegraphics[width=0.85\textwidth]{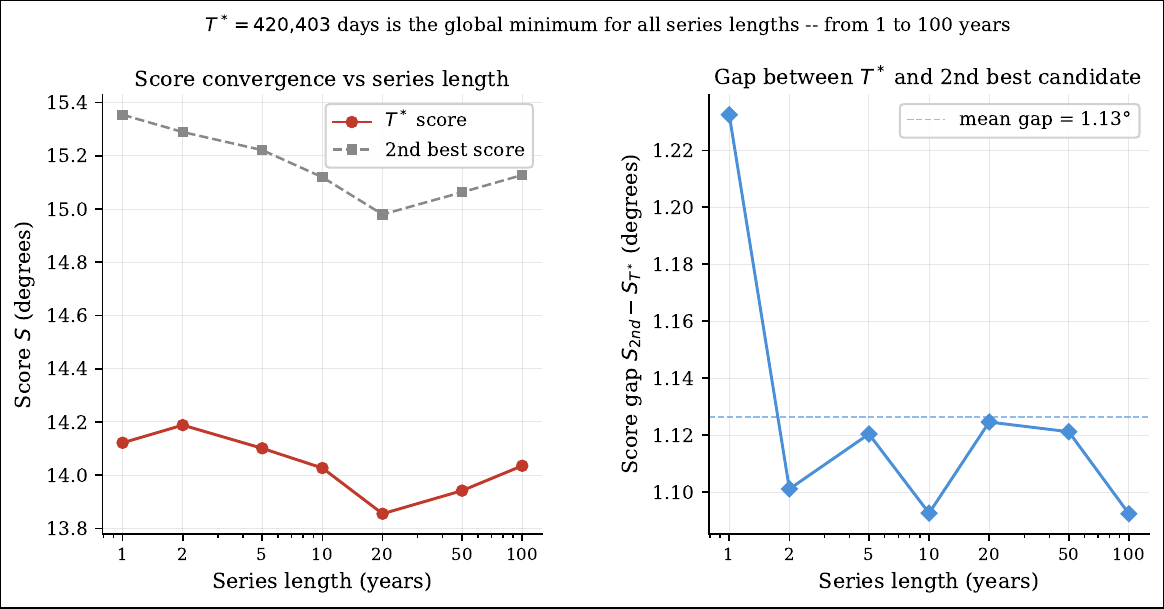}
  \caption{Convergence of the result with series length.
    \emph{Left}: scores of $\Tstar$ (red) and the second-best
    candidate (grey).
    \emph{Right}: gap, stable at $\approx 1.1^\circ$.}
  \label{fig:convergence}
\end{figure}

Table~\ref{tab:convergence} and Figure~\ref{fig:convergence} show the
result for series lengths from 1 to 100~years.
$\Tstar$ is the global minimum for every series length tested, with a
stable gap of $\approx 1.1^\circ$ to the second-best candidate.
The phenomenon is detectable even from a single year of planetary
positions.

\begin{table}[h]
\centering
\caption{Global minimum and score of $\Tstar$ vs series length.}
\label{tab:convergence}
\begin{tabular}{rrrrr}
\toprule
Length (yr) & Best $T$ (yr) & Best score ($^\circ$) &
  $\Tstar$ score ($^\circ$) & Gap ($^\circ$) \\
\midrule
   1 & $-1151$ & 14.122 & 14.122 & 1.232 \\
   2 & $-1151$ & 14.188 & 14.188 & 1.101 \\
   5 & $-1151$ & 14.101 & 14.101 & 1.120 \\
  10 & $-1151$ & 14.027 & 14.027 & 1.093 \\
  20 & $-1151$ & 13.855 & 13.855 & 1.125 \\
  50 & $-1151$ & 13.942 & 13.942 & 1.121 \\
 100 & $-1151$ & 14.036 & 14.036 & 1.092 \\
\bottomrule
\end{tabular}
\end{table}

\subsection{Per-planet breakdown and theoretical residues}
\label{sec:results:perplanet}
\label{sec:results:residues}

Table~\ref{tab:perplanet} shows the mean absolute deviation and signed
mean deviation for each planet at $\Tstar$, together with the
theoretical sidereal residue defined as

\begin{equation}
  r_k = \left\{\frac{\Tstar}{P_k} -
  \left\lfloor\frac{\Tstar}{P_k}\right\rfloor\right\}_{-0.5}^{+0.5}
  \times 360^\circ,
\end{equation}

where $P_k$ is the mean sidereal period and the braces denote
reduction to $(-180^\circ, +180^\circ]$.
Uranus is included for comparison despite not being part of the
primary metric.

\begin{table}[h]
\centering
\caption{Per-planet angular deviations at $\Tstar$ and theoretical
  sidereal residues.
  Uranus is shown for comparison only (not included in the metric).
  Signed mean: positive = planet is systematically ahead of its
  displaced counterpart; negative = behind.}
\label{tab:perplanet}
\begin{tabular}{lrrrr}
\toprule
Planet  & Mean $|\Delta|$ ($^\circ$) & Std $|\Delta|$ ($^\circ$) &
  Signed mean ($^\circ$) & Theor.\ residue ($^\circ$) \\
\midrule
Mercury &  8.49 & 3.07 & $-8.49$  & $-5.44$ \\
Venus   & 19.54 & 0.24 & $-19.54$ & $-19.54$ \\
Earth   &  6.97 & 0.27 & $-6.97$  & $+0.25$ \\
Mars    & 15.02 & 2.61 & $-15.02$ & $-12.19$ \\
Jupiter & 11.71 & 0.60 & $+11.71$ & $+11.79$ \\
Saturn  & 26.70 & 1.96 & $+26.70$ & $+26.55$ \\
Neptune &  5.29 & 0.10 & $-5.29$  & $-5.21$ \\
\midrule
\textit{Uranus} & \textit{---} & \textit{---} &
  \textit{---} & $-108.35$ \\
\midrule
Total (7) & 13.39 & ---  & ---   & 11.57 \\
\bottomrule
\end{tabular}
\end{table}

Venus shows exact agreement between its empirical offset
($-19.54^\circ$) and its theoretical residue ($-19.54^\circ$).
Jupiter and Saturn also agree closely.
Earth has the smallest empirical offset ($6.97^\circ$) and smallest
standard deviation ($0.27^\circ$) among the fast planets, consistent
with $\Tstar$ being almost exactly an integer number of terrestrial
years ($\Tstar/365.25 = 1{,}151.001$, residue $+0.25^\circ$).
Neptune's empirical offset ($-5.29^\circ$) agrees closely with its
theoretical residue ($-5.21^\circ$), and its standard deviation of
$0.10^\circ$ is the lowest of all seven planets, reflecting the
stability of the near-integer relation
$\Tstar / P_\text{Neptune} \approx 6.986$.
The mean absolute theoretical residue ($11.57^\circ$) is consistent
with the empirical mean deviation ($13.39^\circ$), 
confirming that the observed offsets are dominated by the fractional orbital residues of each planet at T*.
In sharp contrast, Uranus's theoretical residue is $-108.35^\circ$,
nearly one-third of a full orbit --- a value incommensurable with all
other planets in the list.

\subsection{Secondary minima}
\label{sec:results:secondary}

Table~\ref{tab:secondary} lists the 10 best candidate intervals.
Sub-multiples of $\Tstar$ ($\approx 355$~yr $\approx 1/3\,\Tstar$,
$\approx 796$~yr $\approx 2/3\,\Tstar$) are significantly worse,
confirming that $\Tstar$ is irreducible.
The third-best interval ($651$~yr $\approx 33 \times P_{JS}$) is
driven by the Jupiter--Saturn conjunction cycle, a distinct and weaker
phenomenon.

\begin{table}[h]
\centering
\caption{Ten best candidate intervals.}
\label{tab:secondary}
\begin{tabular}{rrrrrrl@{}}
\toprule
Rank & $\Delta$yr & $\Delta$days & Mean ($^\circ$) & Std ($^\circ$) &
  Score ($^\circ$) & Relation \\
\midrule
 1 & $-1151$ & $-420{,}403$ & 13.39 & 0.65 & 14.04 & $-T^*$ \\
 2 & $+1151$ & $+420{,}403$ & 14.35 & 0.78 & 15.13 & $+T^*$ \\
 3 & $-651$  & $-237{,}778$ & 32.14 & 1.07 & 33.21 & $\approx 33 P_{JS}$ \\
 4 & $-355$  & $-129{,}664$ & 31.89 & 1.66 & 33.56 & $\approx 1/3\,T^*$ \\
 5 & $-854$  & $-311{,}924$ & 31.30 & 2.54 & 33.84 & $\approx 3/4\,T^*$ \\
 6 & $+651$  & $+237{,}778$ & 32.78 & 1.17 & 33.94 & $\approx 33 P_{JS}$ \\
 7 & $+796$  & $+290{,}739$ & 32.64 & 1.33 & 33.96 & $\approx 2/3\,T^*$ \\
 8 & $-796$  & $-290{,}739$ & 32.71 & 1.31 & 34.02 & $\approx 2/3\,T^*$ \\
 9 & $+500$  & $+182{,}625$ & 33.10 & 1.21 & 34.30 & $\approx 25 P_{JS}$ \\
10 & $-177$  & $-64{,}649$  & 32.91 & 1.49 & 34.40 & $\approx 9 P_{JS}$ \\
\bottomrule
\end{tabular}
\end{table}

The gap between the global minimum ($14.04^\circ$) and the third-best
candidate ($33.21^\circ$) underscores the exceptional nature of
$\Tstar$: it outperforms the next distinct phenomenon by more than
$19^\circ$.

\subsection{Configuration snapshots}

\begin{figure}[h]
  \centering
  \includegraphics[width=0.85\textwidth]{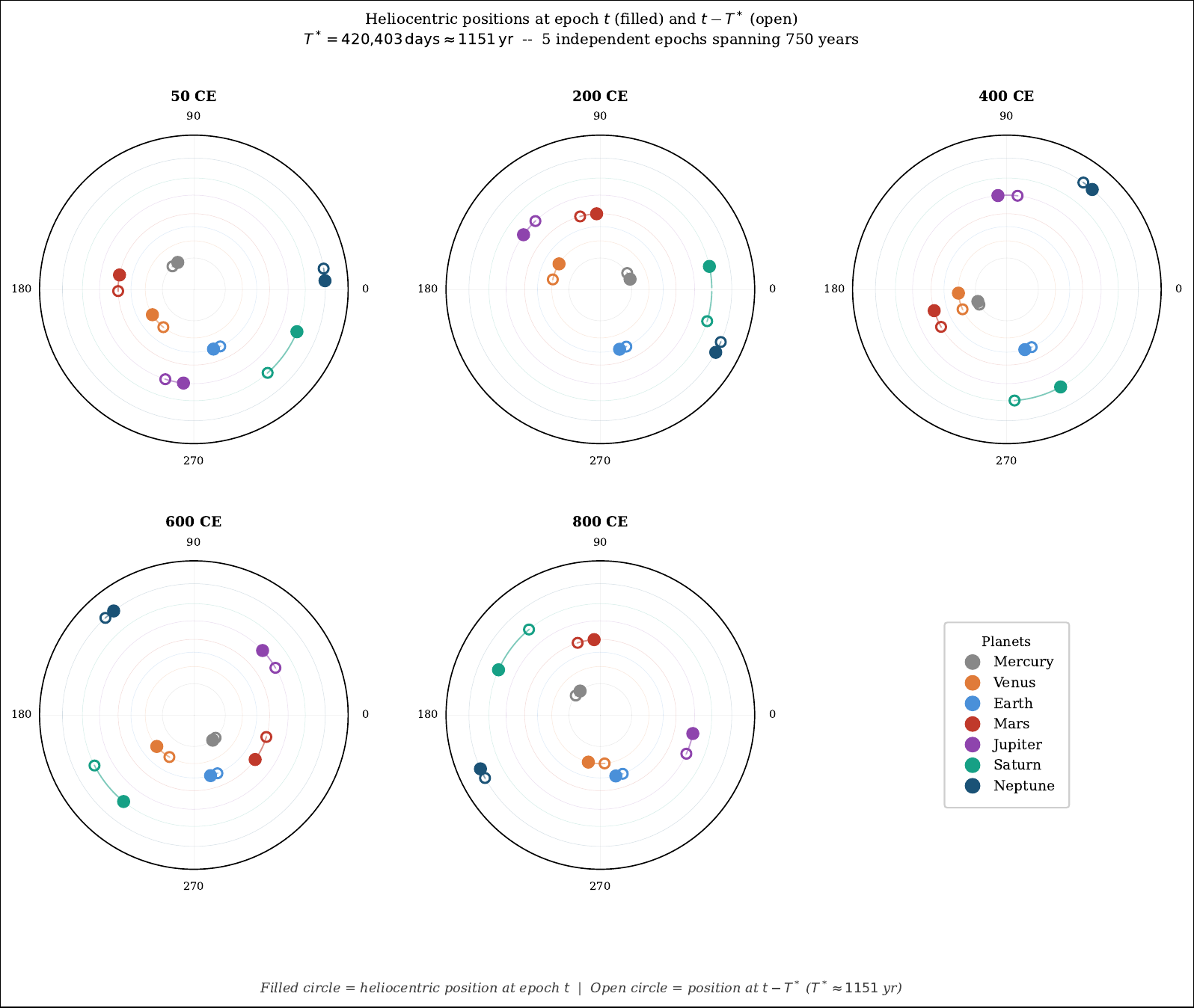}
  \caption{Heliocentric ecliptic positions at epochs 50, 200, 400,
    600, and 800~CE (filled circles) and $\Tstar \approx 1{,}151$~yr
    earlier (open circles).
    Each planet is shown at its normalised orbital radius (Mercury
    innermost, Neptune outermost).
    The near-coincidence persists across all five epochs.}
  \label{fig:polar}
\end{figure}

Figure~\ref{fig:polar} shows heliocentric ecliptic positions of all
seven planets at five independent epochs (50--800~CE), together with
positions $\Tstar$ days earlier.
The near-coincidence of filled and open symbols at every epoch provides
direct visual evidence of the quasi-commensurability and its epoch
independence.

\section{Discussion}
\label{sec:discussion}

\subsection{Arithmetic interpretation}
\label{sec:discussion:arithmetic}

$\Tstar = 420{,}403$~days behaves as an approximate least common
multiple of the seven sidereal periods.
The quality of the approximation varies: Earth is almost exact 
($0.001$ cycles, $+0.25^\circ$); Neptune is precise ($0.014$ cycles, 
$-5.2^\circ$); Mercury is close ($0.015$ cycles, $-5.4^\circ$); 
Venus is intermediate ($0.054$ cycles, $-19.5^\circ$); Mars is 
intermediate ($0.034$ cycles, $-12.2^\circ$); Jupiter and Saturn 
are the least precise ($+11.8^\circ$ and $+26.5^\circ$).

This structure is analogous to, but richer than, classical two-body commensurabilities. The Metonic cycle ($19$~yr) is an approximate LCM of the solar and lunar periods; the Jupiter–Saturn Great Conjunction cycle (~$19.86$ yr) is a near-commensurability of two planetary periods. T* extends this concept to seven bodies simultaneously across a timescale roughly sixty times longer than these classical cycles.

\subsection{The Babylonian Venus period: from one planet to seven}
\label{sec:discussion:babylon}

The interval $1{,}151$~years appears in Babylonian astronomical
records.
\citet{dejong2019} showed that both System~A$_0$ and System~A$_3$ of
Babylonian Venus theory are built on the relation: in $1{,}151$~years,
Venus completes $720$ synodic cycles ($1{,}871$ sidereal orbits),
returning to precisely the same ecliptic longitude.
This relation is equivalent to the near-vanishing of the sidereal
residue of Venus at $\Tstar$, which our computation confirms:
empirical residue $-19.54^\circ$, theoretical residue $-19.54^\circ$
(exact agreement).
The Babylonian astronomers thus identified the sharpest single-planet 
component of the multi-planet quasi-commensurability among the planets 
visible to the naked eye.
The present work demonstrates that this same interval is simultaneously
near-optimal for six additional planets using modern high-precision
ephemerides and an exhaustive search.

\subsection{Geocentric consequences of the quasi-commensurability}
\label{sec:discussion:retrogrades}

Although the analysis is heliocentric, the quasi-commensurability has
direct consequences for the geocentric sky.

A planet's geocentric retrograde motion occurs when its apparent
ecliptic longitude decreases, due to the relative geometry of the
planet and Earth.
Retrograde episodes are kinematic events determined by the instantaneous
angular velocities of both bodies.
Since $\sigma_\delta \approx 0.65^\circ$ over 100~years
(Table~\ref{tab:stability}), the offset between the reference and
candidate series remains nearly constant for centuries, implying that
angular velocities --- and therefore the timing, duration, and extent
of retrograde windows --- are approximately preserved across $\Tstar$.

This was verified directly by identifying all retrograde episodes in
both series and measuring their temporal shifts.
Table~\ref{tab:retrogrades} summarises the results.
For Mercury, the mean shift between corresponding retrograde peaks is
only $+12 \pm 19$~hours across 316~matched episode pairs.
For the outer planets, the shift is larger in absolute terms but
remarkably stable: the standard deviation of the shift is only 12--18~hours for Venus,
Jupiter, and Neptune, and 40~hours for Saturn, meaning each retrograde
episode recurs at a predictable time to within one or two days.
Mars is the exception, with a standard deviation of 204~hours,
consistent with its larger sidereal residue and its sensitivity to
near-commensurabilities with Jupiter.

The shift values are consistent with the sidereal residues of each
planet (Table~\ref{tab:perplanet}): the temporal offset of a retrograde
peak equals approximately the angular residue divided by the planet's
mean angular velocity.

\begin{table}[h]
\centering
\caption{Retrograde episode synchronisation at $\Tstar = 420{,}403$~days.
  For each planet, all retrograde episodes in the 100-year reference
  series are matched to corresponding episodes in the candidate series
  displaced by $\Tstar$, and the shift between retrograde peaks is
  measured.
  A systematic non-zero mean shift reflects the planet's sidereal
  residue (Table~\ref{tab:perplanet}); the standard deviation measures
  the consistency of that shift across all episodes.}
\label{tab:retrogrades}
\begin{tabular}{lrrrr}
\toprule
Planet & Episodes & Mean peak shift & Std & Max \\
       & (matched) & (hours) & (hours) & (hours) \\
\midrule
Mercury &  316 & $+12$  &  19 &   48 \\
Venus   &   63 & $+490$ &  12 &  504 \\
Mars    &   47 & $-417$ & 204 &  840 \\
Jupiter &   91 & $+498$ &  18 &  528 \\
Saturn  &   97 & $+847$ &  40 &  912 \\
Neptune &   99 & $+41$  &  13 &   72 \\
\bottomrule
\end{tabular}
\end{table}

The results are striking for Neptune: despite its slow mean motion,
its retrograde episodes recur with a mean peak shift of only $+41$~hours
and a standard deviation of $13$~hours --- the second smallest of all
six planets, exceeded only by Mercury.
This is a direct consequence of its small sidereal residue ($-5.2^\circ$):
although the angular offset is small, the slow orbital velocity of Neptune
($0.006^\circ$~day$^{-1}$) means that even a modest angular residue
implies a non-negligible temporal shift; the precision of the synchronism
is nonetheless confirmed by the low standard deviation.

In practical terms: an observer at any epoch would find not only a
similar planetary arrangement $1{,}151$~years later, but a similar sky
\emph{evolving similarly} over centuries, with each retrograde episode
recurring at a predictable time to within one or two days for five of
the six planets.
\newpage
\subsection{Effect of the Moon}
\label{sec:discussion:moon}

When the Moon is included as an additional body in a geocentric version
of the analysis, the mean angular deviation at $\Tstar$ increases,
since the Moon's synodic period is $29.531$~days and
$\Tstar / P_\text{Moon} = 14{,}236.19$, leaving a fractional residue
of $0.19$ cycles ($\approx 67^\circ$).
However, the standard deviation of the metric remains low, and
$1{,}151$~years remains the global minimum: the quasi-commensurability
is preserved.
Given the Moon's high angular velocity, a marginally closer
configuration can be found at a short offset from $\Tstar$,
but the cycle length is unchanged.

\subsection{The anomalous exclusion of Uranus}
\label{sec:discussion:uranus}

Seven of the eight planets of the Solar System participate in the
$1{,}151$-year quasi-commensurability.
The sole exception is Uranus.

The contrast with Neptune is striking and deserves emphasis.
Both are ice giants of similar mass and located in the outer Solar
System, yet their participation in the synchronism is entirely
different: Neptune's residue at $\Tstar$ is $-5.2^\circ$, the result
of the near-integer relation
$\Tstar \approx 7 \times P_\text{Neptune}$ (a $0.2\%$ coincidence),
while Uranus's residue is $-108.3^\circ$, nearly one-third of a full
orbit.
No analogous near-integer relation holds for Uranus's orbital period.
When Uranus is added to the seven-planet metric, the score at $\Tstar$
rises sharply, driven entirely by Uranus's large residue.

We note that Uranus is dynamically anomalous among Solar System
planets in a well-documented way.
Its axial tilt of $97.8^\circ$ --- the most extreme of any planet ---
is widely attributed to a giant impact during the early Solar System
\citep{kegerreis2018}, an event that would have substantially altered
its orbital and rotational properties relative to their primordial
values.
\citet{lu2022} showed that Uranus's spin axis precesses too slowly to
be in secular resonance with any relevant frequency of the current
Solar System, placing it dynamically outside the coherent framework
shared by the other planets.

The significance of this is as follows.
If the 1{,}151-year quasi-commensurability reflects a near-integer
arithmetic structure that was present in the Solar System before any
catastrophic perturbations --- as the participation of seven otherwise
diverse planets suggests --- then Uranus's non-participation is
precisely what one would expect if its orbital period was substantially
modified by a giant impact.
The non-participation of Uranus, viewed from this angle, constitutes
an independent empirical line of evidence, derived purely from orbital
periods, that is consistent with the giant-impact hypothesis.

We emphasise that we do not claim to demonstrate the giant-impact
hypothesis by this argument, nor to determine its specific parameters.
The arithmetic near-coincidence identified here is a necessary but not
sufficient condition: Uranus's period could in principle differ from a
near-integer multiple of $\Tstar$ for other reasons.
What we observe is that among eight planets, seven fit the pattern and
one does not, and that the one exception is independently identified as
dynamically anomalous by multiple lines of evidence.
The convergence of these independent observations merits further
theoretical investigation.

If Uranus originally participated in the synchronism, the nearest
integer relation would be $\Tstar \approx 14 \times P_\text{Uranus}$,
implying a pre-impact orbital period of $\approx 82.2$~years and a
semi-major axis of $\approx 18.9$~AU, compared to the current
$84.0$~years and $19.2$~AU.
The dynamical plausibility of this conjecture remains to be tested
against giant-impact simulations \citep{kegerreis2018}.

\subsection{Nature of the phenomenon}
\label{sec:discussion:nature}

The term `resonance' in celestial mechanics denotes a dynamical state
in which bodies exchange angular momentum through gravitational
interaction, maintained by a stabilising mechanism \citep{henrard1997_resonance}.
What we document here is an empirical near-integer relation among
orbital periods --- a \emph{quasi-commensurability} --- whose dynamical
origin we have not investigated.
Whether this arithmetic near-coincidence arises from known resonances
(e.g.\ the near 5:2 commensurability between Jupiter and Saturn),
from Solar System formation history, or from some other cause, lies
beyond the scope of this paper.
Multi-body synchronisms can have dynamical origins: \citet{luque2023}
demonstrated that the six planets of HD~110067 follow a chain of
mean-motion resonances whose architecture has remained essentially
unchanged since the system's formation, showing that such
configurations can be dynamically stable over billions of years.
Whether the quasi-commensurability reported here has a comparable
dynamical foundation is an open question we invite the community to
investigate.

\section{Conclusions}
\label{sec:conclusions}

\begin{enumerate}

  \item \textbf{Global optimum.}
    $\Tstar = 420{,}403$~days ($\approx 1{,}151$~years) is the global
    minimum of $S(T)$ over $2{,}600$ candidates in a $\pm 1{,}300$-year
    symmetric search range, for seven planets (Mercury, Venus, Earth,
    Mars, Jupiter, Saturn, Neptune) using DE441.

  \item \textbf{Statistical exceptionality.}
    Score $14.04^\circ$, rank $1/2{,}600$;
    gap to second best: $1.09^\circ$.

\item \textbf{Temporal stability.}
    Score variance $0.100^\circ$ over a complete cycle of $\Tstar$ 
    ($1{,}210$~years): a structural property of the Solar System, 
    independent of epoch.

  \item \textbf{Robustness.}
    $\Tstar$ is the global minimum for any series length from one year
    upwards, with a stable gap of $\approx 1.1^\circ$ to the second-best
    candidate.

  \item \textbf{Irreducibility.}
    Sub-multiples produce significantly worse scores; the gap to the
    next distinct phenomenon ($651$~yr, consistent with
    $\approx 33 \times P_\text{JS}$) exceeds $19^\circ$.

  \item \textbf{Neptune participates precisely.}
    Neptune's sidereal residue at $\Tstar$ is $-5.2^\circ$, the result
    of the near-integer relation
    $\Tstar \approx 7 \times P_\text{Neptune}$.
    Its empirical offset ($-5.29^\circ$) and standard deviation
    ($0.10^\circ$) confirm it as a precise participant in the
    quasi-commensurability.

  \item \textbf{Uranus is the sole non-participant.}
    Its sidereal residue at $\Tstar$ is $-108.35^\circ$, nearly
    one-third of a full orbit.
    Uranus is the only planet of the Solar System that does not
    participate in the synchronism.

  \item \textbf{Uranus's exclusion as independent evidence.}
    Uranus's non-participation, combined with its independently
    documented dynamical anomalies (extreme axial tilt, anomalous
    spin-axis precession), constitutes an independent empirical
    line of evidence consistent with the hypothesis that its orbital
    period was substantially modified by a giant impact.

  \item \textbf{Consistency with sidereal residues.}
    The empirical mean deviation ($13.39^\circ$) is consistent with
    the mean absolute sidereal residue ($11.57^\circ$), confirming
    that the observed offsets are accounted for by the fractional
    parts of $\Tstar/P_k$ alone.

  \item \textbf{Connection to Babylonian astronomy.}
    The Babylonian $1{,}151$-year Venus period \citep{dejong2019}
    corresponds to the sharpest single-planet component of the
    multi-planet quasi-commensurability, here extended to seven planets.

  \item \textbf{Geocentric consequences.}
    The quasi-commensurability implies synchronisation of planetary
    retrograde motions, with each episode recurring at a predictable
    time (standard deviation 12--40~hours for most planets;
    Table~\ref{tab:retrogrades}).
    Including the Moon degrades but preserves the synchronism.

\end{enumerate}

Complete source code and data are publicly available \citep{code2025}.

\section*{Acknowledgements}
The author thanks Brandon Rhodes for the Skyfield astronomy library
\citep{skyfield}, and acknowledges the use of NumPy \citep{numpy},
Matplotlib \citep{matplotlib}, and the DE441 ephemeris \citep{park2021},
all of which are publicly available.

\bibliographystyle{plainnat}

\end{document}